\documentclass[runningheads]{llncs}

\usepackage{amsmath}
\usepackage{graphicx}
\usepackage{booktabs}
\usepackage{xcolor}
\usepackage{rotating}
\usepackage[hidelinks]{hyperref}
\usepackage{xurl}
\usepackage{microtype}
\usepackage{newunicodechar}
\usepackage{textcomp}
\newunicodechar{Δ}{\ensuremath{\Delta}}
\newunicodechar{α}{\ensuremath{\alpha}}
\newunicodechar{β}{\ensuremath{\beta}}
\newunicodechar{ε}{\ensuremath{\varepsilon}}
\newunicodechar{δ}{\ensuremath{\delta}}
\newunicodechar{√}{\ensuremath{\surd}}
\newunicodechar{×}{\ensuremath{\times}}
\newunicodechar{≥}{\ensuremath{\geq}}
\newunicodechar{≤}{\ensuremath{\leq}}
\newunicodechar{≈}{\ensuremath{\approx}}
\newunicodechar{∈}{\ensuremath{\in}}
\newunicodechar{→}{\ensuremath{\rightarrow}}
\newunicodechar{·}{\ensuremath{\cdot}}

\begin{document}

\title{The Correctness Illusion in\\LLM-Generated GPU Kernels}
\titlerunning{Correctness Illusion in LLM-Generated GPU Kernels}

\author{Dipankar Sarkar\orcidID{0000-0001-5431-6367}}
\authorrunning{D. Sarkar}
\institute{Arizona State University, USA \\
\email{dsarkar3@asu.edu}}

\maketitle

\begin{abstract}
Benchmarks for LLM-generated GPU kernels (KernelBench, TritonBench,
GEAK) score correctness through fixed-shape, small-sample
\texttt{allclose}-style checks. The number of inputs varies between
benchmarks. The shape, dtype, and tolerance are fixed for each kernel.
We test that oracle empirically. We construct a controlled corpus of
24 Triton and CPU stand-in kernels (15 correct controls and 9
LLM-style buggy variants seeded with documented transcription errors)
and re-evaluate it under op-schema-aware seeded fuzzing with a
high-precision (fp64) CPU reference and per-(op, dtype) absolute
tolerances. The seeded oracle flags 9 of 9 buggy kernels and passes 15
of 15 correct controls, at zero precision cost on controls. We extend
the corpus to 26 ops (adding a flash-attention pair) and re-run the
same protocol on five GPU classes (RTX~3060, A10, L40S, A100~SXM4,
H100~NVL). The verdicts are identical across all five GPUs: 10 of 10
illusions caught and 16 of 16 controls clean. The corpus result is
about LLM-style transcription bugs that the
\texttt{allclose}-on-one-shape oracle certifies as correct, not about
the bug rate of any specific deployed LLM. Every flagged failure
replays byte-for-byte from a stored seed.
\end{abstract}

\section{Introduction}
\label{sec:intro}

LLM-generated GPU kernels are now load-bearing. KernelBench
\cite{kernelbench2025}, TritonBench-G (the basis of TritonBench-revised
referenced from~\cite{geak2025}), and GEAK \cite{geak2025} generate
hundreds of CUDA and Triton kernels per evaluation. Agentic systems
such as KernelBand \cite{kernelband2025} and STARK \cite{stark2025}
compose generated kernels into longer pipelines. Every published
benchmark in this family scores correctness through a fixed-shape,
small-sample \texttt{allclose}-style check. KernelBench
\cite{kernelbench2025} draws five random inputs at the reference
shape, for example. The number of samples varies. The shape, dtype,
and tolerance are fixed per kernel.

We test the oracle. We argue, and measure, that it is systematically
optimistic in three specific ways. (i)~The shape candidate set is one
shape per op. Kernels with tail masking, off-by-one accumulation, or
block-size assumptions pass at the chosen shape and fail elsewhere.
(ii)~The dtype candidate set is one dtype per op. fp16 and bf16 are
rarely tested when fp32 is the listed input. Mixed-precision overflow,
underflow, and accumulation errors stay undetected. (iii)~The
tolerance is hand-picked per op. \texttt{atol} and \texttt{rtol} are
typically set one to three orders of magnitude looser than the
kernel's measured error envelope, so loose tolerances absorb real
wrongness.

We construct a controlled corpus of 24 kernels and re-evaluate it
under seeded, op-schema-aware fuzzing with a high-precision (fp64)
reference. 15 of the 24 are correct controls. 9 are LLM-style buggy
variants seeded with documented transcription errors (missing
$0.5\times$ in GELU, \texttt{other=0.0} versus \texttt{-inf} in
softmax tail masking, missing \texttt{sqrt} in RMSNorm, accumulator
overwrite in matmul, missing $1/\sqrt{D}$ in attention, wrong
\texttt{alpha} in LeakyReLU, and three others). The full per-kernel
listing is in Section~\ref{sec:method:corpus}, Table~\ref{tab:corpus}.
The contributions are four.

\begin{itemize}
\item A method. Op-schema-aware shape generation produces per-input
shapes from shared symbolic dims (matmul $A[M,K]\cdot B[K,N]$,
attention $B,H,S,D$). The fuzzer covers the operator's real domain
instead of an arbitrary rank-3 cube.
\item A faithful oracle. The validator compares outputs against an
fp64 CPU reference with per-(op, dtype) absolute tolerances, records
the full element-wise error distribution (max abs, max rel, ULP
percentiles), and detects NaN and Inf for fp16 and bf16. Note that the
validator's tolerance is absolute-only per (op, dtype), in contrast
to PyTorch's \texttt{allclose} which combines an absolute and a
relative term as $|x - y| \leq \texttt{atol} + \texttt{rtol} \cdot
|y|$.
\item A measurement on real GPUs. The daemon-oracle path catches 9 of
9 buggy kernels (four of which are real Triton GPU kernels) on the
single-GPU corpus and passes 15 of 15 correct controls. We then
extend the corpus to 26 ops (adding a flash-attention pair) and
re-run on five GPU classes; the verdicts are identical.
\item A reproducible pipeline. Each failure stores its full input
snapshot in the \texttt{ReproductionInfo}\linebreak[2] payload; a replay script re-runs any
flagged case through the daemon and verifies the verdict matches bit
for bit.
\end{itemize}

\section{Related Work}
\label{sec:related}

\textbf{LLM kernel benchmarks.} KernelBench \cite{kernelbench2025} and
its KernelBench-X extension \cite{kernelbenchx2026} evaluate
LLM-generated CUDA on 250 GPU workloads. Correctness is
\texttt{torch.allclose} against PyTorch reference operators.
TritonBench-G hosts 183 Triton kernels at five difficulty levels;
GEAK \cite{geak2025} introduces TritonBench-revised (a harness-tightened
subset) and a 30-kernel ROCm benchmark. KernelBand \cite{kernelband2025}
and STARK \cite{stark2025} are agentic kernel-optimization systems
built on these benchmarks. None of the five systems vary input shape or
dtype during correctness scoring.

\textbf{DL library fuzzing.} FreeFuzz \cite{freefuzz2022},
DocTer \cite{docter2022}, DeepREL \cite{deeprel2022}, and NablaFuzz
\cite{nablafuzz2023} fuzz the API layer of PyTorch, TensorFlow, JAX,
and OneFlow. Their oracles are differential (cross-backend) or
metamorphic. A 2023 benchmarking study \cite{benchdlfuzzers2023} and its
TOSEM~2025 extension \cite{evalapifuzzers2025} measure that the seven
SOTA API-level fuzzers in this family collectively catch only 6.5\% of
real-world bugs. None of them target the kernel layer (CUDA or Triton
source) where LLM kernel generation operates.

\textbf{GPU memory bugs.} GPU-Fuzz \cite{gpufuzz2026} generates inputs
that probe memory boundary conditions in DL framework CUDA kernels and
uncovered 13 previously-unknown bugs. Its scope is memory safety. Ours
is numerical correctness.

\textbf{Numerical accuracy.} The mixed-precision behaviour of Triton
and PyTorch and its tolerance implications are documented by the
PyTorch team \cite{pytorchmixedprec} and the BF16 study of
Kalamkar et al.~\cite{bfloat16study}. We treat tolerance calibration as
a separate question and address it in the companion paper
\cite{gpuemuP2}.

\section{Method}
\label{sec:method}

\subsection{Op-schema-aware fuzzing}
\label{sec:method:fuzzing}

Each kernel ships a schema. The schema declares shared symbolic dims with
concrete candidate values, plus per-input tensor names that reference
those dims. For matmul: dims $\{M, K, N\}$, inputs $\{a:[M,K], b:[K,N]\}$,
output $[M,N]$. For attention: dims $\{M, N, D\}$, inputs
$\{q:[M,D], k:[N,D], v:[N,D]\}$. The fuzzer samples one value per dim
per test case, deterministically from a master seed through Blake2b
sub-key derivation, then materialises every input. The implementation
lives in \texttt{crates/gpuemu-daemon/src/fuzzer.rs} of the gpuemu
source.

Every dim's candidate set deliberately includes \emph{boundary} values
alongside regular values. For example $H \in \{1, 3, 7, 256, 1025\}$. The
companion paper \cite{gpuemuP3} ablates this choice and shows
that removing the boundary values from the candidate set drops recall
on shape-dependent bugs to zero.

\subsection{Reference oracle}
\label{sec:method:oracle}

For each test case the daemon invokes the op's \texttt{reference}
script. The script is a Python subprocess that reads inputs through
base64-encoded JSON on stdin, computes the operation in fp64, and
casts back to the test dtype. The reference is therefore a
high-precision (fp64) mathematical reference rounded to the target
dtype. It is not necessarily byte-identical to the kernel's chosen
PyTorch or Triton implementation contract, and we do not claim it is.
The reference defines the oracle's notion of ``correct'' for the
purposes of this paper.

The validator compares the kernel's output against the reference with
a per-(op, dtype) absolute tolerance. We use absolute-only thresholds
per (op, dtype) deliberately, in contrast to PyTorch's
\texttt{allclose} which combines an absolute and a relative term as
$|x - y| \leq \texttt{atol} + \texttt{rtol} \cdot |y|$. The companion
paper \cite{gpuemuP2} measures the calibration of these absolute
thresholds against the kernel's observed error envelope.

The validator also runs NaN and Inf detection extended to fp16 and
bf16, and records the full element-wise \texttt{ErrorStats}
distribution per case: count, num\_exceeding, max abs, mean abs, p50
abs, p90 abs, p99 abs, max rel, mean rel, max ULP, mean ULP. The ULP
distribution becomes the empirical basis for the tolerance
calibration in \cite{gpuemuP2}.

\subsection{The corpus}
\label{sec:method:corpus}

The single-GPU evaluation runs on 24 ops: 15 correct controls and 9
LLM-style buggy variants. We extend the corpus to 26 ops by adding a
flash-attention control plus its buggy variant for the cross-GPU
sweep in Section~\ref{sec:eval:cross}. Table~\ref{tab:corpus} lists
the full canonical corpus. Each row gives the kernel name, the
implementation (numpy stand-in or Triton), the role (correct control
or LLM-style buggy variant), the dtypes covered, and the bug pattern
each buggy variant encodes.

\begin{table}[h]
\centering
\caption{Canonical corpus. 16 controls and 10 LLM-style buggy variants
across 26 kernel entries. The single-GPU evaluation
(Section~\ref{sec:eval}) runs on rows 1--24; the cross-GPU sweep
(Section~\ref{sec:eval:cross}) extends to rows 25--26 (flash-attention
pair).}
\label{tab:corpus}
\resizebox{\textwidth}{!}{%
\begin{tabular}{rlllll}
\toprule
\# & kernel name & impl. & role & dtypes & bug encoded (buggy rows only) \\
\midrule
 1 & \texttt{softmax}                    & numpy  & control & fp32, fp16 & --- \\
 2 & \texttt{layernorm}                  & numpy  & control & fp32, fp16 & --- \\
 3 & \texttt{matmul}                     & numpy  & control & fp32, fp16 & --- \\
 4 & \texttt{softmax\_llm\_buggy}        & numpy  & buggy   & fp32, fp16 & tail-mask leak \\
 5 & \texttt{softmax\_triton}            & triton & control & fp32, fp16 & --- \\
 6 & \texttt{softmax\_triton\_buggy}     & triton & buggy   & fp32, fp16 & \texttt{other=0.0} instead of \texttt{-inf} \\
 7 & \texttt{gelu\_triton}               & triton & control & fp32, fp16 & --- \\
 8 & \texttt{gelu\_triton\_buggy}        & triton & buggy   & fp32, fp16 & dropped leading $0.5$ \\
 9 & \texttt{silu\_triton}               & triton & control & fp32, fp16 & --- \\
10 & \texttt{silu\_triton\_buggy}        & triton & buggy   & fp32, fp16 & \texttt{sigmoid(2x)} ($\beta$ confusion) \\
11 & \texttt{rmsnorm\_triton}            & triton & control & fp32, fp16 & --- \\
12 & \texttt{rmsnorm\_triton\_buggy}     & triton & buggy   & fp32, fp16 & forgot \texttt{sqrt} \\
13 & \texttt{l2norm\_triton}             & triton & control & fp32, fp16 & --- \\
14 & \texttt{l2norm\_triton\_buggy}      & triton & buggy   & fp32, fp16 & forgot \texttt{sqrt} \\
15 & \texttt{relu\_triton}               & triton & control & fp32, fp16 & --- \\
16 & \texttt{leaky\_relu\_triton}        & triton & control & fp32, fp16 & --- \\
17 & \texttt{leaky\_relu\_triton\_buggy} & triton & buggy   & fp32, fp16 & $\alpha = 0.1$ instead of $0.01$ \\
18 & \texttt{sigmoid\_triton}            & triton & control & fp32, fp16 & --- \\
19 & \texttt{tanh\_triton}               & triton & control & fp32, fp16 & --- \\
20 & \texttt{elu\_triton}                & triton & control & fp32, fp16 & --- \\
21 & \texttt{matmul\_triton}             & triton & control & fp32, fp16 & --- \\
22 & \texttt{matmul\_triton\_buggy}      & triton & buggy   & fp32, fp16 & \texttt{acc=} instead of \texttt{acc+=} \\
23 & \texttt{attention\_triton}          & triton & control & fp32, fp16 & --- \\
24 & \texttt{attention\_triton\_buggy}   & triton & buggy   & fp32, fp16 & dropped $1/\sqrt{D}$ score scale \\
\midrule
\multicolumn{6}{l}{\textit{Cross-GPU extension (Section~\ref{sec:eval:cross}):}} \\
25 & \texttt{flash\_attention\_triton}        & triton & control & fp32, fp16 & --- \\
26 & \texttt{flash\_attention\_triton\_buggy} & triton & buggy   & fp32, fp16 & dropped $\text{acc}\cdot\alpha$ rescale after max update \\
\bottomrule
\end{tabular}%
}
\end{table}

\subsection{Pipeline}
\label{sec:method:pipeline}

The experimental harness provisions an ephemeral GPU instance on
vast.ai, labelled for safe teardown, builds the daemon from source,
installs the Python client and Triton, runs the P1 driver against the
corpus, and uploads results to Backblaze B2 under a run id. Teardown
is guaranteed by a context-manager \texttt{\_\_exit\_\_}, with a
label-strict reaper that cleans up orphan instances. The full
artefact details are in the public corpus repository (see
Section~\ref{sec:conclusion}).

\subsection{Assumptions}
\label{sec:method:assumptions}

The empirical claim that follows depends on five assumptions, which we
state plainly so readers can audit them.

\begin{enumerate}
\item The 10 buggy variants (9 in the single-GPU corpus plus a flash-
attention variant in the cross-GPU extension) are author-seeded with
documented LLM transcription patterns. They are not pulled directly
from real LLM outputs. The result is therefore about which bug
patterns the \texttt{allclose}-on-one-shape oracle certifies as
correct, not about the bug rate of any specific deployed LLM.
Section~\ref{sec:limitations} discusses why we accept this trade.
\item The high-precision (fp64) reference, rounded to the target
dtype, defines the oracle's notion of ``correct'' for the purposes of
this paper. The reference is not guaranteed to be byte-identical to
any particular library's correctness contract. Treating fp64 round-off
as zero is standard practice in mixed-precision validation
\cite{pytorchmixedprec,bfloat16study}.
\item Per-(op, dtype) absolute tolerances are set ahead of time and
held fixed during the P1 evaluation. The companion
paper~\cite{gpuemuP2} addresses whether those tolerances are
themselves correct.
\item Triton kernels are compiled fresh per run. We do not control for
ptxas randomness across providers. The cross-GPU result in
Section~\ref{sec:eval:cross} indicates this does not affect the
correctness verdict.
\item The Python client decodes received tensors as contiguous, so
non-contiguous layout variation is nominal at the client boundary even
when the daemon-side fuzzer varies strides.
\end{enumerate}

\section{Evaluation}
\label{sec:eval}

\textbf{Setup.} The primary single-GPU run uses an RTX~3060 instance
on vast.ai with image
\texttt{pytorch/pytorch:2.4.0-cuda12.4-cudnn9-devel}, 30 iterations
per (op, dtype) for the 24 single-GPU ops and 2 dtypes, about 1{,}440
cases. The run id on Backblaze B2 is
\texttt{run-20260611-095210-889b18}.

\textbf{P1 headline.} Across the 24-op single-GPU corpus, the seeded
oracle flags 9 of 9 LLM-style buggy variants and passes 15 of 15
correct controls. There are no false positives on controls.
Table~\ref{tab:verdicts} reports the verdict per kernel for a 12-op
subset; the full per-kernel result is in
\texttt{summary.json} alongside the run id above and in the corpus
repository.

\begin{table}[h]
\centering
\caption{P1 verdict per kernel on the 24-op single-GPU corpus, 12-op
subset. \emph{bench} is the fixed-shape allclose-style oracle.
\emph{gpuemu} is the seeded oracle. An \emph{illusion} is a kernel
the bench oracle passes but the seeded oracle fails.}
\label{tab:verdicts}
\resizebox{\textwidth}{!}{%
\begin{tabular}{lllllc}
\toprule
kernel & source & bench & gpuemu & fail/total & illusion \\
\midrule
gelu\_triton\_buggy        & llm:demo-triton & pass & \textbf{fail} & 29/30 & YES \\
rmsnorm\_triton\_buggy     & llm:demo-triton & pass & \textbf{fail} & 30/30 & YES \\
silu\_triton\_buggy        & llm:demo-triton & pass & \textbf{fail} & 29/30 & YES \\
softmax\_llm\_buggy        & llm:demo        & pass & \textbf{fail} & 10/30 & YES \\
softmax\_triton\_buggy     & llm:demo-triton & pass & \textbf{fail} & 13/30 & YES \\
l2norm\_triton\_buggy      & llm:demo-triton & pass & \textbf{fail} & 30/30 & YES \\
leaky\_relu\_triton\_buggy & llm:demo-triton & pass & \textbf{fail} & 30/30 & YES \\
matmul\_triton\_buggy      & llm:demo-triton & pass & \textbf{fail} & (K$>$1) & YES \\
attention\_triton\_buggy   & llm:demo-triton & pass & \textbf{fail} & (D$\geq$8) & YES \\
gelu\_triton               & human:triton    & pass & pass          & 0/30 & --- \\
softmax\_triton            & human:triton    & pass & pass          & 0/30 & --- \\
matmul\_triton             & human:triton    & pass & pass          & 0/30 & --- \\
\bottomrule
\end{tabular}%
}
\end{table}

\begin{figure}[h]
\centering
\includegraphics[width=\textwidth]{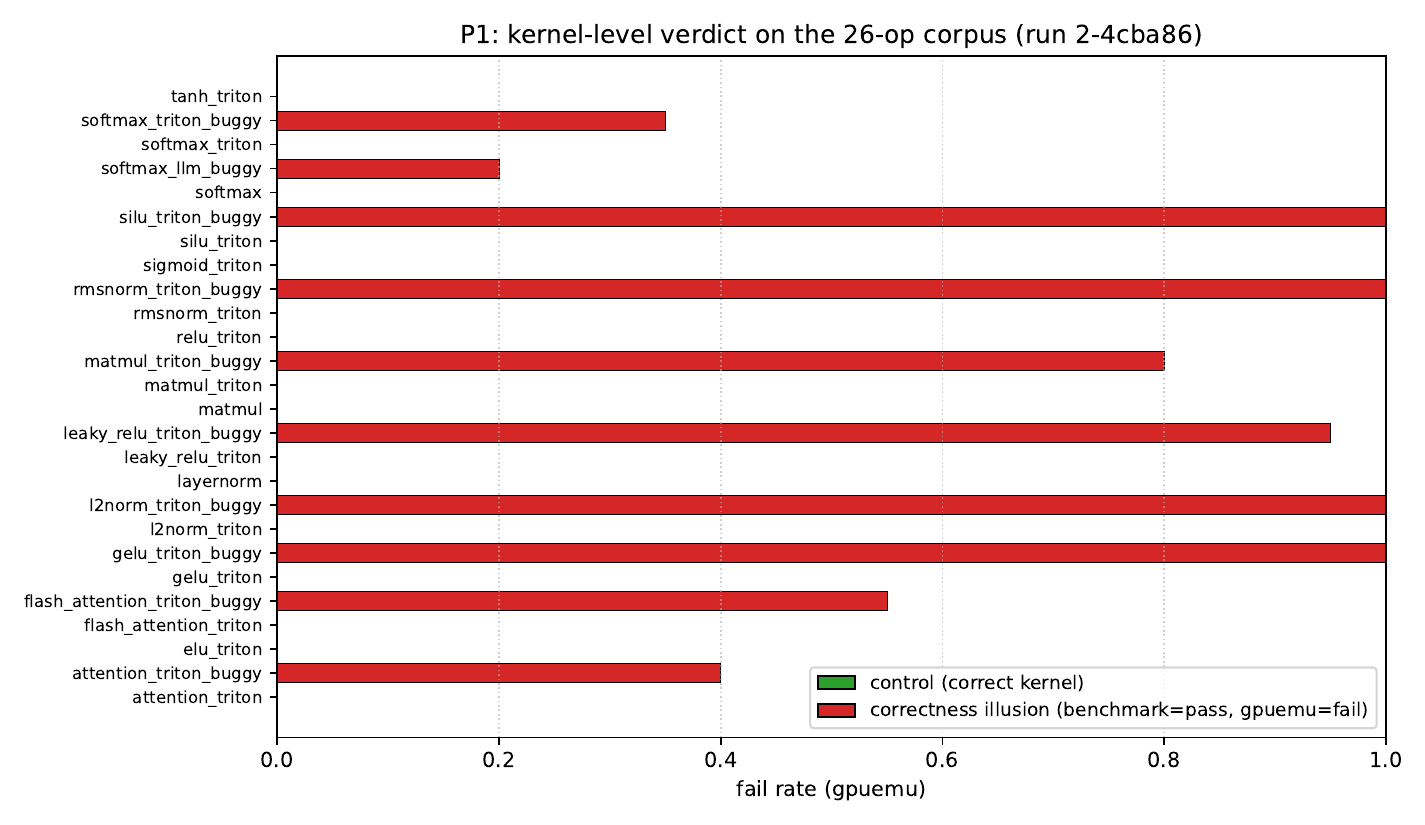}
\caption{Verdict per kernel on the full 26-op corpus, plotted from
the RTX~3060 cross-GPU run. Green indicates correct controls that
pass cleanly. Red indicates illusions (bench oracle pass, seeded
oracle fail). The cross-GPU sweep in §4.1 confirms the same verdict
on the four remaining GPU classes.}
\label{fig:verdicts}
\end{figure}

The tail-mask family (\texttt{softmax\_*}, \texttt{leaky\_relu})
shows shape-dependent illusion. \texttt{softmax\_triton\_buggy} fails
13 of 30 randomly sampled cases. Under a regular-shapes-only strategy
it fails 0 of 10. Under a boundary-only strategy it fails 6 of 10.
The companion paper~\cite{gpuemuP3} ablates this pattern across seven
input-generation strategies on the same corpus. The bug surfaces at
$H = 3$ (tail less than \texttt{BLOCK}) and vanishes at $H = 256$
(power of two, no tail).

\textbf{Minimal failing cases.} Each illusion ships a minimal failing
shape through the daemon's \texttt{Minimize} endpoint. Examples:
\texttt{softmax\_triton\_buggy} minimises to $[1, 1, 3]$ fp16;
\texttt{gelu\_triton\_buggy} minimises to $[2, 1, 1]$ fp16. The
allclose-style oracle would not exercise either shape.

\textbf{Reproducibility.} Each failure stores a binary input
snapshot. A replay script fetches the run from the result store,
decodes the snapshot, and replays the kernel through the daemon. The
verdict matches bit for bit.

\subsection{Cross-architecture consistency}
\label{sec:eval:cross}

We extend the corpus to 26 ops by adding a flash-attention control
and its LLM-style buggy variant (rows 25--26 in
Table~\ref{tab:corpus}). We then re-run the same protocol on five GPU
classes: RTX~3060 (sm\_86), A10 (sm\_86), A100~SXM4 (sm\_80), L40S
(sm\_89), and H100~NVL (sm\_90). The verdicts are identical across
all five GPUs: 16 of 16 controls pass cleanly, and 10 of 10
LLM-style illusions are caught. The 10 illusions include the real
Triton matmul, attention, and flash-attention buggy variants.

\begin{figure}[h]
\centering
\includegraphics[width=\textwidth]{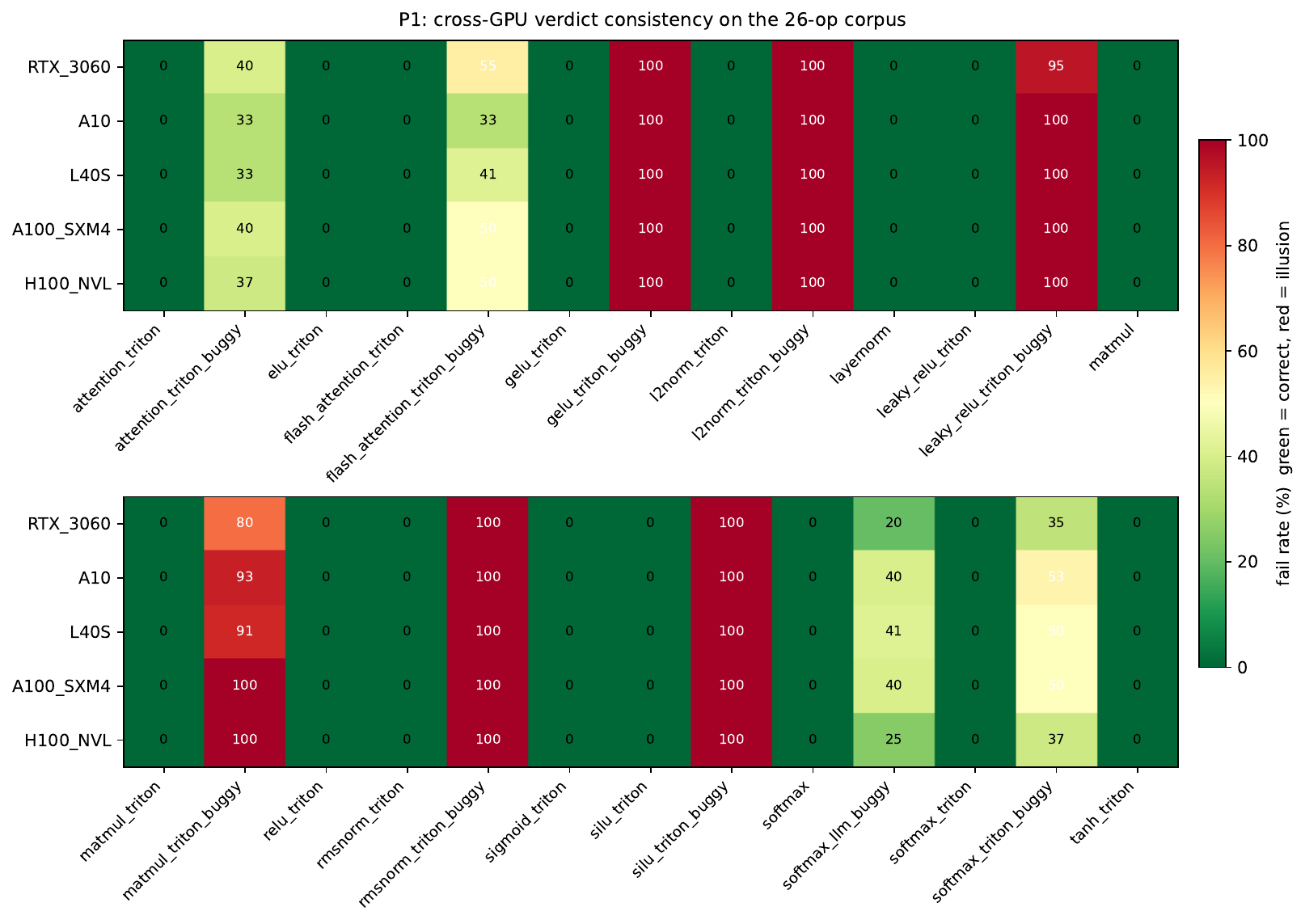}
\caption{Cross-GPU verdict consistency on the 26-op corpus. Each
panel covers half the corpus; rows are the five GPU classes; cells
are per-kernel fail rates. Controls stay green on every GPU.
Illusions stay red on every GPU.}
\label{fig:crossgpu}
\end{figure}

The harness's safety contract (label-strict reaper, confirmed
destroy, context-manager teardown) prevented orphan instances across
the five parallel launches even when two providers returned
\texttt{Bind for 0.0.0.0:18433 failed: port is already allocated}
during provisioning. Figure~\ref{fig:crossgpu} reports the per-kernel
verdicts as two stacked panels (kernels 1--13 on top, 14--26 on
bottom) so the per-cell labels stay readable at textwidth.

\section{Discussion}
\label{sec:discussion}

The bug categories gpuemu surfaces partition cleanly into two classes.

\textit{Magnitude-uniform bugs} (gelu missing $0.5$, silu
\texttt{sigmoid(2x)}, leaky\_relu wrong $\alpha$, rmsnorm and l2norm
missing \texttt{sqrt}, attention missing the score scale) are caught
on essentially every shape. The LLM benchmark missed them only because
it tested one shape and the absolute tolerance was set above the
constant bias the bug introduces.

\textit{Shape-dependent bugs} (softmax tail mask, matmul \texttt{acc=}
instead of \texttt{acc+=}) are visible only when the schema includes
the right boundary shape. Without boundary-aware fuzzing they are
invisible. The companion paper \cite{gpuemuP3} measures this gap.

The two classes call for different countermeasures. The first wants
tighter tolerances \cite{gpuemuP2}. The second wants better input
generation \cite{gpuemuP3}.

\section{Limitations}
\label{sec:limitations}

The 10 buggy variants are author-seeded with documented LLM
transcription bugs. They are not pulled from real LLM-generated
kernel outputs. The result is therefore about which bug patterns the
\texttt{allclose}-on-one-shape oracle certifies as correct, not about
the bug rate of any specific deployed LLM. We choose author-seeded
for two reasons. First, ground truth is exact. Second, the eight bug
families we seed are documented in the LLM-Triton literature and
represent a deliberate threat model rather than a sample of
convenience. A natural extension is to fuzz LLM-generated kernels
from GEAK~\cite{geak2025} or KernelBench~\cite{kernelbench2025}
directly.

The validator currently does same-dtype comparison: kernel-fp16
against reference-fp16 rounded from fp64. Cross-dtype comparison
(kernel-fp16 against reference-fp64) is a noted future extension.

The Python client decodes received tensors as contiguous, so
non-contiguous layout fuzzing is nominal at the client boundary even
when the daemon-side fuzzer varies strides.

bfloat16 is supported on the daemon protocol but the Python client
lacks a native bf16 dtype (a NumPy limitation). We proxy to fp16 in
tests.

\section{Conclusion}
\label{sec:conclusion}

LLM-style transcription bugs in GPU kernels can pass a fixed-shape,
small-sample \texttt{allclose}-style oracle as ``correct''. The
op-schema-aware seeded oracle in this paper exposes the gap. On the
24-op single-GPU corpus, every LLM-style buggy variant is caught and
every correct kernel passes. The extended 26-op cross-GPU sweep
replays the same verdict on five GPU classes. The countermeasures
(boundary-aware shape sets, per-(op, dtype) tolerance
calibration~\cite{gpuemuP2}, principled input
generation~\cite{gpuemuP3}) are within reach of any project that
already runs an allclose-style oracle today.

\paragraph{Artefact.}
The 26-op corpus, the driver that ran this experiment, the analysis
script that produced every table, and the replay tool that fetches
each cited run record from object storage are bundled in the public
\textsf{gpuemu-corpus} package at
\url{https://github.com/sarkar-dipankar/gpuemu-corpus}. The validator
daemon is at \url{https://github.com/Skelf-Research/gpuemu}. Each
flagged failure replays byte for byte from a stored seed.

\paragraph{License.}
This preprint is released under
\href{https://creativecommons.org/licenses/by/4.0/}{CC-BY 4.0}.

\bibliographystyle{splncs04}
\bibliography{refs}

\end{document}